\documentclass[preprint,showpacs,preprintnumbers,superscriptaddress,amsmath,amssymb]{revtex4}

\usepackage[dvips]{epsfig}
\usepackage{graphicx}
\usepackage{dcolumn}
\usepackage{bm}
\usepackage[english,german]{babel}
\usepackage{amsmath}
\usepackage{amssymb}
\usepackage[latin1]{inputenc}

\newcommand{\mat}[1]{\boldsymbol{#1}}

\begin{document}


\title{Solid-solid phase transition in hard ellipsoids}

\author{M. Radu}
\email{schillit@uni-mainz.de}
\affiliation{\selectlanguage{german}
  Institut f"ur Physik, Johannes Gutenberg-Universit"at, Staudinger Weg 7, D-55099
  Mainz, Germany
}
\author{P. Pfleiderer}
\affiliation{
  Department of Chemical Engineering, K.U. Leuven, W. de Croylaan 46,
  B-3001 Leuven, Belgium
}
\author{T. Schilling}
\email{schillit@uni-mainz.de}
\affiliation{\selectlanguage{german}
  Institut f"ur Physik, Johannes Gutenberg-Universit"at, Staudinger Weg 7, D-55099
  Mainz, Germany
}

\selectlanguage{english}

\date{\today}

\begin{abstract}
We present a computer simulation study of the crystalline phases of hard
ellipsoids of revolution. A previous study [Phys.~Rev.~E, \textbf{75}, 
020402 (2007)] showed that for aspect ratios
$a/b\ge 3$ the previously suggested stretched-fcc phase 
[Mol.~Phys., \textbf{55}, 1171 (1985)] is unstable with respect to
a simple monoclinic phase with two ellipsoids of different
orientations per unit cell (SM2). In order to study the stability of these
crystalline phases at different aspect ratios and
as a function of density we have calculated their
free energies by thermodynamic integration. The integration path was sampled 
by an expanded ensemble method in which the weights were adjusted by the 
Wang-Landau algorithm.
 We show that for aspect ratios $a/b\ge 2.0$ the SM2 structure is
more stable than the stretched-fcc structure for all densities above
solid-nematic coexistence. Between $a/b=1.55$ and $a/b=2.0$ our calculations 
reveal a solid-solid phase transition. 
\end{abstract}

\pacs{05.20.Gg,05.70.Ce,02.70.Rr,82.70.Dd,61.50.Ah,82.20.Wt,65.40.gd}

\maketitle

\section{\label{sec:Intro}Introduction}
Suspensions of hard particles (i.e. particles that interact via an infinitely 
strong, repulsive excluded-volume interaction potential) have been
successfully used as model systems for the statistical mechanics of liquids 
and solids for more than half a century. 
For this class of system phase transitions are entropy rather than
enthalpy driven, and the relevant control parameters are the particle 
shape and concentration rather than temperature. 
Hard ellipsoids are a simple model for systems whose macroscopic properties
depend on the interplay of positional and orientational entropy such as 
liquid crystals \cite{frenkel.mulder:1985,
zarragoicoechea.levesque.weis:1992, allen.mason:1995, 
camp.mason.allen.khare.kofke:1996}
and orientational glasses \cite{letz.schilling.latz:2000,
  demichele.schillingr.sciortino:2007, pfleidererschilling08}

In recent years it has been shown by computer simulations and experiments 
that randomly packed arrangements of hard ellipsoids can reach
densities much higher than random close packing of spheres
\cite{DonevScience04, Sacanna07, bezrukov07}. At certain aspect
ratios, random packing of ellipsoids can even reach densities almost as 
high as the closest crystalline packing of spheres \cite{DonevScience04}. 
However, this does not imply that random packing of ellipsoids is as dense 
as their densest known crystalline packing. In 2004, Donev and co-workers
introduced a family of crystalline packings of ellipsoids, which reach 
a packing fraction of $\eta \simeq 0.7707$ \cite{donev04}  
(as compared to $\eta = \frac{\pi}{\sqrt{18}} \simeq 0.7405$ for the fcc packing of spheres
and stacking variants thereof). 

\begin{figure}[h]
  \centerline{\psfig{figure=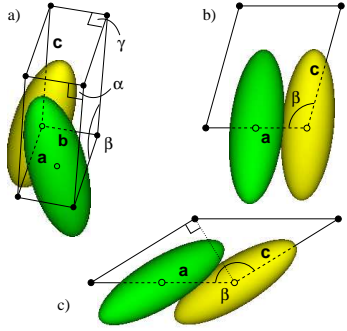,width=45mm}}
  \caption{Unit cell of SM2 \cite{pfleidererschilling07} with $a/b = 3$ (color
    online). The open circles indicate the centers of the two ellipsoids
    which form the basis. The yellow (light gray) ellipsoid is at the
    origin, the green (dark gray) one is at $\frac{1}{2}({\bf a}+{\bf
      b})$. The cell is monoclinic. $\beta$ is the soft degree of
    freedom. Part c) shows the cell at close packing (the
    infinite-pressure limit), where it is an instance of the family of
    packings introduced by Donev \emph{et al.}\ \cite{donev04}. Note the
    indicated right angle and the resulting symmetry about the
   $bc$-plane in this case. 
   \label{fig:Intro1}}
\end{figure}

Inspired by this study, we re-examined the phase
diagram of hard ellipsoids \cite{pfleidererschilling07}. We found that the
stretched fcc-phase, which had before been assumed to be the stable 
crystalline phase \cite{frenkelmulder02}, was unstable with respect to a 
different crystalline phase. The more stable structure has a simple 
monoclinic unit cell containing two ellipsoids of unequal orientation
(SM2)(cf.~Fig \ref{fig:Intro1}). The packings constructed by Donev and
co-workers \cite{donev04} are a special case of SM2 (the
infinite-pressure limit).

At that time we did, however, not compute free energy differences between 
SM2 and stretched-fcc. In the present article we report on Monte Carlo 
simulations in which SM2 and stretched-fcc are connected to their respective 
harmonic crystals (``Einstein crystals'')
via thermodynamic integration, and hence their free energies are determined.
In order to sample the thermodynamic integration pathway, we adapted
the Wang-Landau algorithm \cite{wanglandau01}. In the original 
Wang-Landau scheme a flat histogram
of the internal energy is constructed. Here we constructed a flat histogram of 
the coupling parameter that couples the hard ellipsoid model to the
Einstein crystal, instead.

\section{\label{sec:CompMeth}Method of Computation}

In order to determine which of two phases is thermodynamically more stable, 
one compares their relevant thermodynamic potentials, e.g.~in the case of
constant particle number $N$, volume $V$ and temperature $T$ their 
free energies $F$. Within a MC simulation, however, for most models it is
impossible to compute $F$ because of its direct connection to
the accessible phase space volume ($\mat{q}^N$,$\mat{p}^N$). To solve this 
problem the method of Thermodynamic Integration (TI)
\cite{frenkelladd84, frenkelsmit02} is commonly used, in which the 
free energy difference between the system of interest and a reference 
system can be calculated by introducing an artificial external potential 
$U$, such that
\begin{equation}
    \Delta F = F_{\rm sys}-F_{\rm ref} = \int_{\zeta = 1}^{\zeta = 0}\mathrm{d}\zeta\,\bigg\langle\frac{\partial
      U(\mat{q}^N;\zeta)}{\partial\zeta}\bigg\rangle_{\zeta}\,\,.\label{eq:TI1}
\end{equation}
Here, the parameter $\zeta\in [0,1]$ links the
interaction potential of the system of interest $U_{\rm sys}\equiv U(\zeta=0)$
to the potential of the reference
system $U_{\rm ref}\equiv U(\zeta=1)$ by
\begin{equation}
  U(\mat{q}^N;\zeta) = (1-\zeta)U_{\rm sys}(\mat{q}^N) + \zeta U_{\rm ref}(\mat{q}^N)\,\,.\label{eq:TI2}
\end{equation}

During a typical integration, $U_{\rm sys}$ is gradually switched on and
at the same time $U_{\rm ref}$ is gradually switched off. However, the
hard-core interaction of the ellipsoids does not allow for a gradual
change. Therefore, $U_{\rm sys}$ is imposed in a first step,
and then $U_{\rm ref}$ is gradually switched off in a second. With this procedure,
the free energy of the system can be calculated as $F_{\rm sys} =
F_{\rm ref} + \Delta F_1 + \Delta F_2$, where the subscripts refer to the two
steps just described. 
\begin{equation}
  \Delta F_1 = -\ln\langle\exp[-\beta U_{\rm sys}]\rangle_{\zeta = 1} \,,\label{eq:TI3}
\end{equation} 
where $U_{\rm sys}(\mat{q}^N)$ is here the overlap potential of the ellipsoids, 
and the configuration $\mat{q}^N$ consists of 
positions and orientations $\mat{q}^N\equiv\lbrace\mat{r}^N,\theta^N\rbrace$. 
$\langle\ldots\rangle_{\zeta}$ refers to the ensemble average where the
potential is parameterized by Eq.~\ref{eq:TI2}. 
$\Delta F_2$ will be discussed together with the free energy of 
the reference state in the following paragraph.

\begin{figure}[h]
  \centerline{\psfig{figure=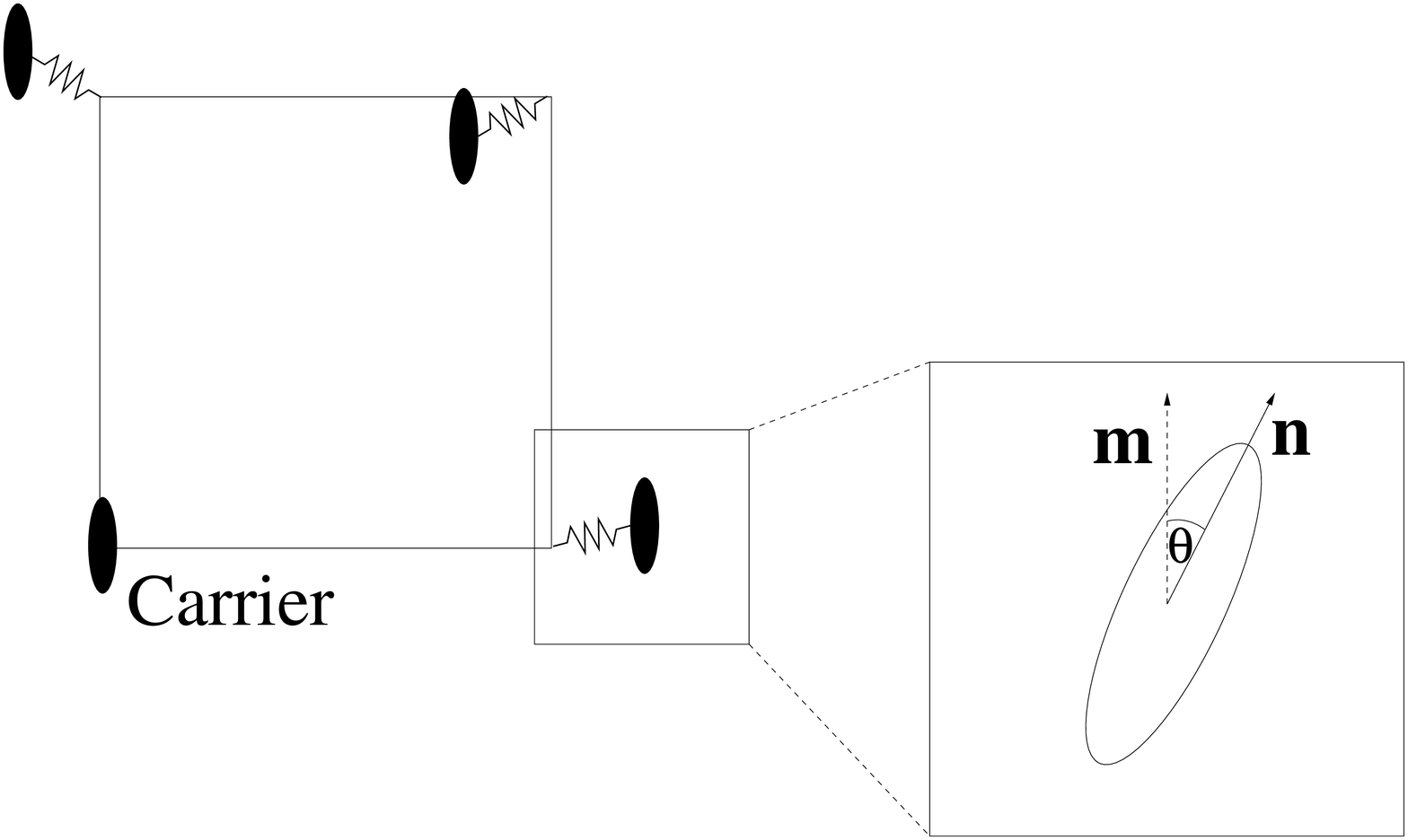,width=70mm}}
  \caption{Scheme of an Einstein Molecule for hard ellipsoids}
  \label{fig:EM1}
\end{figure}

From Eq.~\ref{eq:TI1} it is obvious that $F_{\rm ref}$ needs to be known from
other sources, e.g.~by analytical computation and that no phase transition
may occur during the integration process.
In order to construct such a reference system, we consider a 
system of hard ellipsoids in which all particles except for
one are coupled to the sites of a lattice via harmonic springs. The
remaining particle is fixed in space and is called the carrier of the
lattice. We fix this particle to the origin of the coordinate system.
As we are interested in anisotropic particles, we will also 
restrict their rotational motion by a contribution 
$U_{\rm rot}(\theta_i^N)$ to the
potential. We set this to be proportional to $\sin^2\theta_i$, where $\theta_i$
is the angle between the axis of particle $\mat{n}_i$ and a reference axis
$\mat{m}_i$ (cf.~Fig \ref{fig:EM1}).

This kind of model is known as an Einstein Molecule (EM)
\cite{veganoya07}. (The reason for fixing one particle is the following: 
In the case of an Einstein Crystal (EC), center of mass motion of the entire system 
does not cost energy. Hence, for weak coupling one needs to shift all 
particle positions after every move to keep the center of mass positioned, 
as it was done e.g.~in the work of Polson \emph{et al.} \cite{polson99}. 
In the case of the EM, the fixed carrier particle
ensures non-divergency of the center of mass mean square displacement 
for a negligible harmonic potential (see also \cite{veganoya07}).) 

The interaction potential of the EM is 
\begin{align}
  U_{\rm ref} &= \sum_i\!^\prime\,\lambda_i^{\rm trans}(\mat{r}_i
  -\mat{r}_{0,i})^2 + \sum_i\!^\prime\,\lambda_i^{\rm rot}\sin^2\theta_i \nonumber \\
  &= \lambda \sum_i\!^\prime\,\Big[(\mat{r}_i-\mat{r}_{0,i})^2 + \sin^2\theta_i
  \Big] \,\,,\label{eq:EM1}
\end{align}
where $\mat{r}_{0,i}$ denote the position vectors of the lattice
sites. The prime denotes that the sum runs over all particles except for the
carrier. For simplicity we chose the spring constants of all lattice
sites in the second line of Eq.~\ref{eq:EM1} as equal. In addition we
set $\lambda^{\rm trans} = \lambda^{\rm rot} \equiv \lambda$.
We use twice the short axis $b$ as the
unit of length and $k_BT$ as the unit of energy (except where stated
otherwise). With this the unit of $\lambda$ is $k_BT/(2b)^2$. As we
are only interested in the configurational part of phase space,
the kinetic energies of the particles are disregarded.\\
In order to evaluate the configurational part of the partition function 
of the Einstein Molecule, we 
assume that the maximum coupling constant $\lambda_{\rm max}$ is strong enough 
for $\theta_i \ll 1$. So we obtain (cf. \cite{veganoya08})
\begin{align} 
  \frac{F^{\rm FCC}}{N} &= \frac{1}{N}\ln\bigg[\frac{N}{4\pi V}\bigg] +
  \frac{3}{2}\bigg(1-\frac{1}{N}\bigg)\,\ln\bigg[\frac{\lambda_{\rm max}}{\pi}\bigg]\nonumber\\
  &\qquad + \bigg(1-\frac{1}{N}\bigg)\,\ln\bigg[\frac{\lambda_{\rm max}}{2\pi}\bigg]\,\,. \label{eq:EM2}
\end{align}
In case of the SM2-EM the same approach leads to
\begin{align}
  \frac{F^{\rm SM2}}{N} &= \frac{1}{N}\ln\bigg[\frac{N}{8\pi V}\bigg] +
  \frac{3}{2}\bigg(1-\frac{1}{N}\bigg)\,\ln\bigg[\frac{\lambda_{\rm max}}{\pi}\bigg]\nonumber\\
  &\qquad + \bigg(1-\frac{1}{N}\bigg)\,\ln\bigg[\frac{\lambda_{\rm max}}{2\pi}\bigg]\,\,. \label{eq:EM3}
\end{align}
The derivation of Eq.~\ref{eq:EM3} is outlined in Appendix
\ref{app:AppA}. The difference in free energy per
particle between the FCC-EM and the SM2-EM is $(\ln 2)/N$, due to the 
presence of two types of lattice sites in the SM2
unit cell. This difference vanishes in the thermodynamic limit $N \to
\infty$.\\
Coming back to the calculation of $F_{\rm sys}$ we rewrite the
integral in Eq.~\ref{eq:TI1} (with $\zeta = \lambda/\lambda_{\rm max}$) as
\begin{equation}
\Delta F_2 =
  \int_{\lambda_{\rm max}}^{0}\mathrm{d}\lambda\,\Bigg[\Big\langle\sum_i\!^\prime\,
 (\mat{r}_i - \mat{r}_{0,i})^2\Big\rangle_{\lambda} + \Big\langle\sum_i\!^\prime\,
  \sin^2\theta_i\,\Big\rangle_{\lambda}\Bigg] \,\, . \label{eq:EM4}
\end{equation}

We evaluate eqs.~\ref{eq:TI3} and \ref{eq:EM4} by the following 
expanded ensemble technique: We discretize the range of values for 
$\lambda$. Then, besides translational and
rotational moves, we perform a move in which the system passes
from one model with a value $\lambda_i$ into an adjacent one with $\lambda_j$
and vice versa. In order to ensure good statistics when sampling the
$\lambda$-range, we introduce a set of weights $\psi_m$ and sample the
expanded ensemble given by the partition function
\begin{equation}
  \mathcal{Z} =
  \sum_{m=1}^M\mathcal{Z}_m(\lambda_m)\,e^{\psi_m}\,\,, \label{eq:EE2}
\end{equation}
where $\mathcal{Z}_m(\lambda_m)$ is the partition function of the 
model $m$ with $\lambda = \lambda_m$ and $\psi_m$ its weighting factor. 
The acceptance probability of a $\lambda$-move is
then given by 
\begin{equation}
  P_{i\to j} =
  \rm{min}\bigg[1,\frac{\exp(\psi_j-U_j)}{\exp(\psi_i-U_i)}\bigg]\,\,, \label{eq:EE3}
\end{equation}
such that for an adequate set of weights the system can be forced to
visit the states of interest. One can then compute the free energy
difference as
\begin{equation}
  F_j - F_i = -\ln\bigg[\frac{p_j}{p_i}\bigg] + \psi_j -
  \psi_i\,\,. \label{eq:EE4}
\end{equation}
Here $p_i$ and $p_j$ are the probabilities for the system to visit model $i$
or model $j$, respectively, in the presence of the weights.

The success of this procedure depends on finding appropriate weights. 
The weights are not 
known \emph{a priori}, but they can be adjusted iteratively during the 
simulation, as has been introduced by Wang and Landau \cite{wanglandau01} 
for the case of the density of states as a function of energy. We apply 
this idea to thermodynamic integration. 
Initially, we choose the weighting factors as
$\psi_i = 0\,\,\,\forall\,i\in\lbrace 0,\ldots,M\rbrace$. 
Then simulations for each subensemble are carried out in
which after each $\lambda$-move the weight of the rejected model 
is increased by $\Delta\psi = 1$. 
This leads to an increase of both the possibility to visit the
accepted model and the possibility to stay there. As in the
original algorithm $\Delta\psi$ is decreased by a factor $a < 1$,
$\Delta\psi\to a\cdot\Delta\psi$ (here: $a=0.5$), as soon as the
difference between the probabilities becomes sufficiently
small. Once the simulations are finished 
$p_m\approx p_{m-1}$ (in fact $\ln[p_m/p_{m-1}]$ was less than $10^{-5}\,Nk_BT$ after
only $10^5$ steps).

Finally, we consider the
computation of $\Delta F_1 = F_{\rm off} - F_{\rm on}$, where the
$\lambda$-step does not change $\lambda$ but consists of switching on
and off the hard-core potential.

According to Eq.~\ref{eq:EE3} moves which switch off the potential
or which lead to a state with no overlap are always accepted whereas
moves of the form off $\to$ on which yield a state with at least two
overlaping particles are always rejected.
For this case the coupling parameter was $\lambda = \lambda_{\rm max}$
(i.e.~the reference state), and hence the free energy difference
between the states on and off was expected to be very small.

Therefore we set the corresponding weights equal to $0$ and kept them
fixed during the calculation. (This approach is validated by our
results for $\Delta F_1$, which were of the order of $10^{-4}\,Nk_BT$.)

\section{\label{sec:HEll}Results}
\subsection{\label{subsec:HS}Hard spheres}
\begin{figure}[h]
  \centerline{\psfig{figure=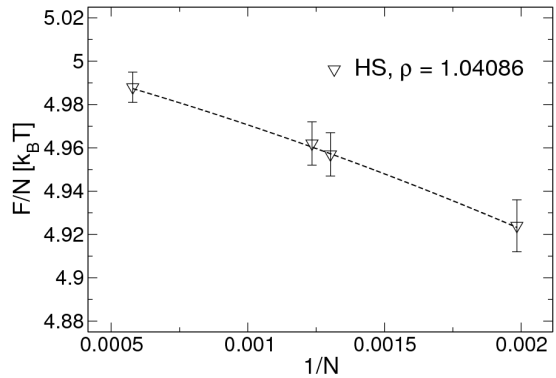,width=72mm}}
  \caption{Hard spheres: Free energy per particle as a function of the inverse 
particle number. Symbols: MC data. Line: Fit according to eq.~\ref{eq:HS2}}
  \label{fig:HS1}
\end{figure}

In order to test the algorithm before applying it to anisotropic
particles, we first computed the free energy of hard spheres at various 
densities $\varrho=N/V$ and particle numbers $N$. 
Table \ref{tab:HS1} summarizes 
our results. Fig.~\ref{fig:HS1} shows the free energy per particle as a 
function of $1/N$ for $\varrho=1.04086$. The dotted 
line is a fit to
\begin{equation}
  \frac{F}{N} = e_1 + \frac{e_2}{N} + \frac{e_3}{N^2}
  \,\, ,  \label{eq:HS2}
\end{equation}
by which we extrapolate our results to infinte $N$ (see also 
ref.~\cite{veganoya07}).

\vspace{5mm}

\begin{figure}[h]
  \centerline{\psfig{figure=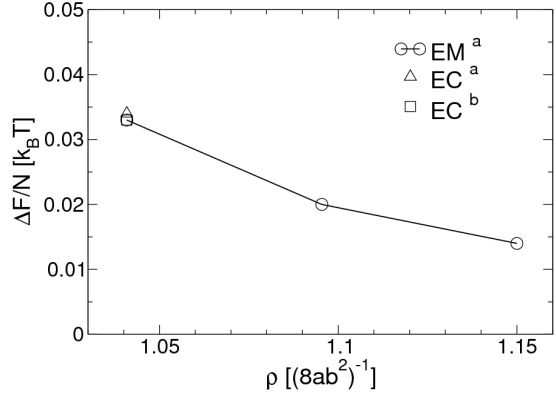,width=72mm}}
  \caption{Hard spheres: Difference $\Delta F$ between our results and free 
    energies computed by similar methods ((a) ref.~\cite{veganoya07} and
    (b) ref.~\cite{polson99}). There is very good agreement.
  \label{fig:HS2}}
\end{figure}

Fig.~\ref{fig:HS2} shows a comparison of our results with Einstein Crystal and 
Einstein Molecule computations that did not use
the Wang Landau algorithm. There is very good agreement, giving us confidence
in the results for the hard ellipsoid case.
\begin{table}[h]
  \caption{\label{tab:HS1}Results for the free
    energy of hard spheres}
  \begin{ruledtabular}
    \begin{tabular}{ccc}
      $\varrho$ & $N$ & $F/N$ \\
      \hline
      $1.04086$ & $504$ & $4.924(12)$ \\
      & $768$ & $4.957(10)$ \\
      & $810$ & $4.962(10)$ \\
      & $1728$ & $4.988(7)$ \\
      $1.09975$ & $1728$ & $5.647(7)$ \\
      $1.15000$ & $1728$ & $6.283(7)$ \\
    \end{tabular}
  \end{ruledtabular}
\end{table}

\subsection{Hard ellipsoids}
Previous work showed that the angle of inclination of the SM2 unit 
cell, $\beta$, is a very ``soft'' degree of 
freedom \cite{pfleidererschilling07}, i.e.~the corresponding shear 
modulus is almost zero. $\beta$ fluctuated
strongly even at a pressure as high as $P=46\; {\rm k_BT}/{8ab^2}$ (for 
$a/b=3$, where the nematic-solid coexistence pressure 
is $P=31\; {\rm k_BT}/{8ab^2}$ \cite{frenkel.mulder:1985}). This unusual
mechanical property is due to the fact that planes of equally oriented
particles can slide across each other without much interaction, unless the
system is forced to pack very densely. In order to quantify this effect, we
computed free energies for various fixed values of $\beta$.
In the special case that the unit cell is invariant under reflections with 
respect to the bc-plane (see Fig.~\ref{fig:Intro1}), the configuration 
has the same symmetry as
(but different unit cell parameters than) the close-packed
structure constructed by Donev \emph{et al.} \cite{donev04}. In the 
following we refer to this structure as $SM2^{\rm (cp)}$. 

\begin{figure}[h]
  \centerline{\psfig{figure=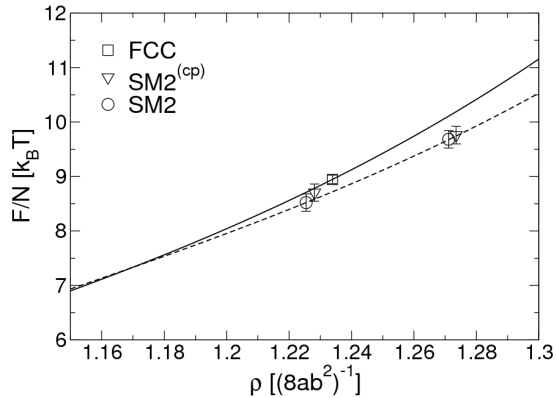,width=72mm}}
  \caption{Hard ellipsoids, $a/b=3$: Free energy per particle 
    as a function of density. Symbols are direct MC results for the free 
    energy, lines are fits to MC data for the equation of state.}
  \label{fig:HEllDen1}
\end{figure}

Fig.~\ref{fig:HEllDen1} shows the free energy for $a/b = 3$
as a function of the density $\varrho$ (to simplify the comparison with 
other studies we use $1/8ab^2$ as the unit of density here instead of
$1/8b^3$). The symbols are direct simulation results of free energies: 
triangles for $SM2^{\rm (cp)}$, 
circles for SM2 with different values of $\beta$ 
(see table \ref{tab:HEllDen1} 
for details) and a square for stretched fcc. The
solid lines are polynomial fits to the equation of state data from 
our previous work (ref.~\cite{pfleidererschilling07}), integrated 
over $\varrho$ and shifted by a constant to fit the free energy data.

Taking the errors into account there is no evidence for a difference in 
free energy between the different angles of inclination $\beta$ for the 
SM2 crystals.
This supports our earlier observation that the angle of inclination is a soft
degree of freedom \cite{pfleidererschilling07}.

For decreasing density, the free energy difference between stretched fcc 
and SM2 decreases and the lines intersect 
at $\varrho \simeq 1.17$, which is very close to the solid-nematic phase
transition ($\varrho = 1.163$ according to
ref.~\cite{frenkel.mulder:1985}). Our data therefore confirm that
SM2 is more stable than fcc at $a/b = 3$ and above $\varrho \simeq 1.17$.

\begin{table}[h]
  \caption{\label{tab:HEllDen1}Hard Ellipsoids: Free energy per
    particle for $a/b=3$}
  \begin{ruledtabular}
    \begin{tabular}{ccccc}
      Lattice type & $\varrho$ & $N$ & $\beta\,[^\circ]$ & $F_{\rm sim}/N$ \\
      \hline
      FCC & $1.23390$ & $432$ & $(90)$ & $8.95(8)$ \\
      $\rm SM2$ & $1.22545$ & $432$ & $110.76$ & $8.52(16)$ \\
      & $1.27111$ & $432$ & $115.91$ & $9.68(16)$ \\
      $\rm SM2^{\rm (cp)}$ & $1.22815$ & $432$ & $148.35$ & $8.71(16)$ \\
      & $1.27352$ & $432$ & $147.97$ & $9.76(16)$ \\
    \end{tabular}
  \end{ruledtabular}
\end{table}

\begin{figure}[h]
  \centerline{\psfig{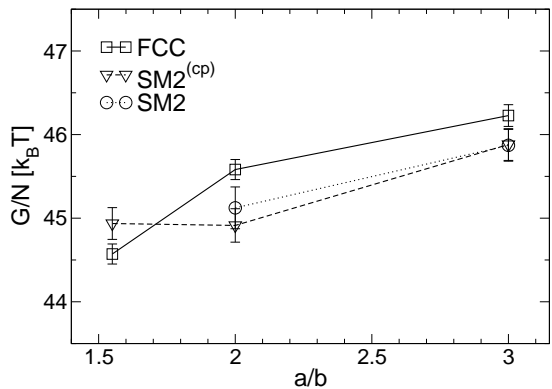}}
  \caption{Hard ellipsoids: Gibbs free energy per particle vs. aspect ratio $a/b$ at 
$P=46\,k_BT/8ab^2$. Lines to guide the eye. At $a/b\ge 2$, SM2 is more stable, 
while fcc is more stable at $a/b=1.55$, implying a solid-solid phase 
transition in between.}
  \label{fig:HEllAspRat2}
\end{figure}

In Table \ref{tab:HEllAspRat1}, we compare the free energies of SM2,
$SM2^{\rm (cp)}$ and fcc as a function of aspect ratio, viz.~for $a/b
= 1.55,2$ and $3$. 
As our input configurations were produced at a fixed pressure 
($P=46\,k_BT/8ab^2$), systems of different aspect ratios and/or
structure had different densities. In order to compare them, we calculated 
the Gibbs free energy per particle $G/N$ by the Legendre transform of $F/N$ 
with respect to the volume.  
Fig.~\ref{fig:HEllAspRat2} shows the Gibbs free energy per particle. (The
lines are guides to the eye, only.)
Again there is no difference (within the errorbars) 
between the $SM2^{\rm (cp)}$ structure and the other values of
$\beta$. The superior stability of SM2 is confirmed for $a/b \ge
2$. At $a/b = 1.55$, however, we find that fcc is more
stable, indicating a phase transition
between $a/b = 1.55$ and $a/b=2.0$. 

(This happens to be near $a/b = \sqrt{3}$ , the lower
boundary of aspect ratios for which prolate
ellipsoids can form crystals with maximal packing fraction $\eta =
0.770732$ \cite{donev04}; but smaller aspect ratios near this value
still produce higher-than-fcc densities, so that we do not suspect a
connection.)\\

\begin{table}[h]
  \caption{\label{tab:HEllAspRat1}Hard ellipsoids: Free energy and Gibbs free
    energy per particle ($P=46\,k_BT/8ab^2$)}
  \begin{ruledtabular}
    \begin{tabular}{ccccccc}
      Lattice type & $a/b$ & $\varrho$ & $N$ & $\beta\,[^\circ]$ & $F/N$ & $G/N$\\
      \hline
      FCC & $1.55$ & $1.23027$ & $1728$ & $(90)$ & $7.18(7)$ & $44.57(12)$ \\
          & $2.00$ & $1.23171$ & $1728$ & $(90)$ & $8.24(7)$ & $45.58(12)$\\
          & $3.00$ & $1.23390$ & $432$ & $(90)$ & $8.95(8)$ & $46.23(13)$\\
      $\rm SM2$ & $2.00$ & $1.27215$ & $432$ & $142.44$ & $8.96(23)$ & $45.12(25)$\\
                 & $3.00$ & $1.27111$ & $432$ & $115.91$ & $9.68(16)$ & $45.87(19)$\\
      $\rm SM2^{\rm (cp)}$ & $1.55$ & $1.25013$ & $768$ & $127.42$ & $8.14(16)$ & $44.94(19)$\\
                  & $2.00$ & $1.27702$ & $432$ & $135$ & $8.89(17)$ & $44.91(20)$\\
                  & $3.00$ & $1.27352$ & $432$ & $147.97$ & $9.76(16)$ & $45.88(19)$\\
    \end{tabular}
  \end{ruledtabular}
\end{table}

\begin{figure}[h]
  \centerline{\psfig{figure=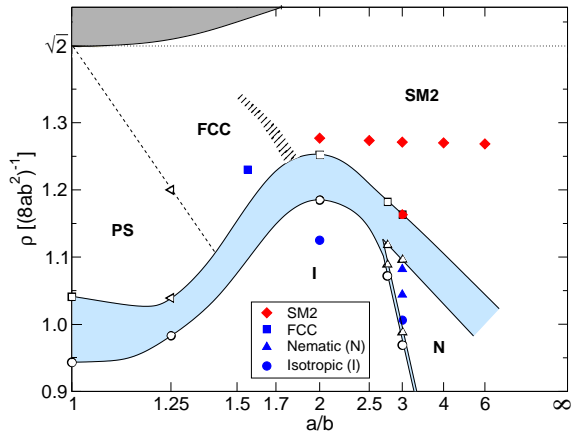,width=75mm}}
  \caption{(color online). Updated phase diagram of hard ellipsoids of
    revolution. It includes part of the results of Frenkel and Mulder \cite{frenkel.mulder:1985}
    (open symbols), and their suggested phase boundaries and coexistence
    regions. The data points at $a/b = 1$ are taken from
    \cite{hooverree68}. We have inserted our state points (this work
    and \cite{pfleidererschilling08}; filled symbols), and extended
    its high-density boundary to the maximum densities found by Donev
    \emph{et al.}~\cite{donev04}, hence including all densities
    possible in SM2 (recall that SM2 at maximum packing coincides with
    the packings of Donev \emph{et al.}~). In hashes we indicate a
    possible location of the coexistence region between fcc and SM2
    (see text for details).
  }
  \label{fig:HEllAspRat3}
\end{figure}

In Fig.~\ref{fig:HEllAspRat3} we show an updated phase diagram of hard
ellipsoids of revolution. It includes part of the results of Frenkel
and Mulder \cite{frenkel.mulder:1985}, and their suggested phase
boundaries and coexistence regions. We have inserted our state points
(this work and \cite{pfleidererschilling08}), and extended its
high-density boundary to the maximum densities found by Donev \emph{et
  al.}~\cite{donev04}, hence including all densities possible in SM2
(recall that SM2 at maximum packing coincides with the packings of
Donev \emph{et al.}~). As stated above, our data imply a phase
transition between SM2 and fcc near $a/b=\sqrt{3}$. 

In hashes we
indicate a possible location of the coexistence region, according to
the following argument: For spheres ($a/b=1$) and
maximum packing ($\varrho=\sqrt{2}$) the density differences
among plastic solid, fcc and SM2 vanish, so the coexistence regions
among these phases should join and vanish in width either at this point, or 
before this point is
reached. (We can not make statements yet about the details of the
approach to the sphere limit, therefore we have not drawn anything) . 
For $a/b>1$, and above $\varrho=\sqrt{2}$ (the dotted line), only SM2 exists, 
so the SM2-fcc coexistence region must lie below $\varrho=\sqrt{2}$. The 
packing efficiency of SM2 and the resulting entropic advantage should 
favor SM2 even below $\varrho=\sqrt{2}$, and the stronger this advantage, 
the lower the transition density --
hence the downward slope of the coexistence region with increasing $a/b$. 
The packing advantage also dictates the increase in width of the region, 
since SM2 is accordingly higher in density at a given pressure. Finally, the
coexistence region should pass between our state points of fcc at
$a/b=1.55$, and SM2 at $a/b=2.00$. The width there we estimate from
the density difference at $a/b=1.55$ and $a/b=2.00$ at pressure
$P=46\,k_BT/8ab^2$ (Table \ref{tab:HEllAspRat1}).

\section{\label{sec:Concl}Conclusion}
In summary, we have studied crystalline phases of hard
ellipsoids considering their relative
stability. We calculated the absolute free energies as
functions of the particle density $\varrho$ and the aspect ratio
$a/b$ by use of a thermodynamic integration technique with an 
Einstein Molecule as the reference state. The integration path was sampled 
by an expanded ensemble method in which the weights were adjusted by the 
Wang-Landau algorithm. After checking our
simulations for reliability considering the test case of hard spheres, we
applied our methods to ellipsoids. At pressure $P=46\,k_BT/8ab^2$ we found 
no difference in the free energies of SM2 crystals with different angles of
inclination $\beta$ . Furthermore our results show that the SM2 phase is more
stable than the stretched-fcc phase for densities $\varrho \gtrsim 1.17$ 
(at $a/b=3$) and for aspect ratios $a/b \ge 2.0$ 
(at $P=46\,k_BT/8ab^2$). Hard ellipsoids exhibit a fcc-SM2 phase transition
between $a/b = 1.55$ and $a/b = 2.0$.

\begin{acknowledgments}
We thank D. Frenkel, M. P. Allen, A. Donev, and W. A. Siebel for helpful
discussions.
We are grateful to the DFG (Tr6/D5 and Emmy Noether Program) for financial
support and to the NIC J\"ulich for CPU time on the JUMP. 
\end{acknowledgments}

\appendix

\section{\label{app:AppA}Calculation of $F^{\rm SM2}$}
First, without loss of generality, we label particle $i=1$ as
the carrier of the lattice. Then we write down the partition function
of the SM2-EM using Eq.~\ref{eq:EM1}:
\begin{align}
  \mathcal{Z}^{\rm SM2} &=
  \Gamma^{(N)} \times \int\mathrm{d}\mat{r}_1\int\mathrm{d}\Omega_1\nonumber\\
  &\qquad \times
  \int\mathrm{d}\mat{r}_2\ldots\mathrm{d}\mat{r}_N\,\exp\Bigg[-\lambda_{\rm
      max}\sum_{i = 2}^N(\mat{r}_i-\mat{r}_{0,i})^2\,\Bigg] \nonumber \\
  &\qquad \times
  \int\mathrm{d}\Omega_2\ldots\mathrm{d}\Omega_N\,\exp\Bigg[-\lambda_{\rm
      max}\sum_{i = 2}^N\sin^2\theta_i\,\,\Bigg]\,\,. \label{eq:A1}
\end{align}
The two trivial integrations are due to our freedom of
choosing $\mat{r}_1$ as origin of the coordinate system and
$\theta_1$ as some orientation in space.

\begin{figure}[h]
  \centerline{\psfig{figure=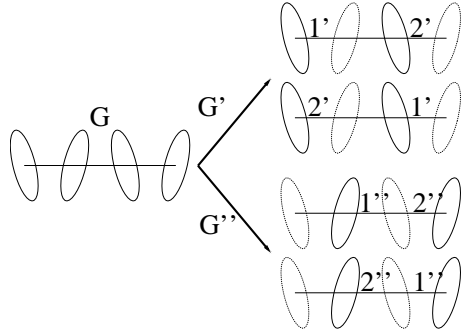,width=60mm}}
  \caption{Combinatorics for an Einstein Molecule of the SM2 type}
  \label{fig:A1}
\end{figure}

$\Gamma^{(N)}$ is a combinatorial factor: We consider a lattice
$G$ which consists of $N$ particles with two different orientations
(distinguished by the primes in Fig.~\ref{fig:A1}). We now divide $G$
up into two sublattices $G^\prime$ and $G^{\prime\prime}$ with respect
to the particle types. On these sublattices there are $(N/2)!$
possibilities, respectively, to position the particles on their
sites. To account for the presence of the carrier on one of the
sublattices the associated factorial is $(N/2-1)!$. Hence
\begin{equation}
  \Gamma^{(N)} = \frac{\big(\frac{N}{2}\big)!\,\big(\frac{N}{2}-1\big)!}{\big(\frac{N}{2}\big)!\,
    \big(\frac{N}{2}\big)!} = \frac{2}{N} \,\, . \label{eq:A2}
\end{equation}

The integral over the spatial coordinates can directly be carried out
and leads to $(\pi / \lambda_{\rm max})^{(3(N-1)/2)}$. Hence, Eq.~\ref{eq:A1} 
can be simplified to

\begin{align}
  \mathcal{Z}^{\rm SM2} &\propto
  \int\mathrm{d}\theta_2^{\prime}\ldots\mathrm{d}\theta_{n_1}^{\prime}\,\theta_i^{\prime}\,\exp\Bigg[-\lambda_{\rm
      max}\sum_{i = 2}^{n_1}\theta_i^{\prime\,2}\,\,\Bigg]\nonumber\\
  &\qquad \times
  \int\mathrm{d}\theta_{n_1}^{\prime\prime}\ldots\mathrm{d}\theta_N^{\prime\prime}\,\theta_j^{\prime\prime}\,\exp\Bigg[-\lambda_{\rm
      max}\sum_{j = n_1}^N\theta_j^{\prime\prime\,2}\,\,\Bigg]\,\,. \label{eq:A3}
\end{align}

Here, because of the azimuthal symmetry of the problem, we already
carried out the integration over $\phi$ and used the approximation
$\sin\theta\approx\theta$ which was motivated before. The two
remaining integrals now are related to the sublattices
$G^{\prime}$ and $G^{\prime\prime}$. Solving them, we find
for the resulting partition function
\begin{equation}
  \mathcal{Z}^{\rm SM2} = \frac{8\pi
    V}{N}\bigg(\frac{\pi}{\lambda_{\rm
      max}}\bigg)^{\frac{3(N-1)}{2}}\,\bigg(\frac{2\pi}{\lambda_{\rm max}}\bigg)^{N-1}\,\,.\label{eq:A4}
\end{equation}


\end{document}